\documentclass[a4paper,12pt]{article}

\usepackage{amssymb,latexsym}

\usepackage{fullpage}

\makeatletter \@addtoreset{equation}{section} 
\makeatother

\usepackage{graphicx} 
\begin{document} 
\begin{titlepage}
	\thispagestyle{empty} 
	\begin{flushright}
		\hfill{hep-th/0702088} 
	\end{flushright}
	
	\vspace{35pt} 
	\begin{center}
		{ \LARGE{\bf Flow Equations for \\[4mm]
		Non-BPS Extremal Black Holes }}
		
		\vspace{60pt}
		
		{Anna Ceresole$^\star$ and Gianguido Dall'Agata$^\dagger$ }
		
		\vspace{30pt}
		
		{\it $\star$ INFN, Sezione di Torino $\&$ Dipartimento di Fisica Teorica\\
		Universit\`a di Torino, Via Pietro Giuria 1, 10125 Torino, Italy}
		
		\vspace{30pt}
		
		{\it $\dagger$ Dipartimento di Fisica ``Galileo Galilei'' $\&$ INFN, Sezione di Padova \\
		Universit\`a di Padova, Via Marzolo 8, 35131 Padova, Italy}
		
		\vspace{40pt}
		
		{ABSTRACT} 
	\end{center}
	
	\vspace{10pt}
	
	We exploit some common features of black hole and domain wall solutions of (super)gravity theories coupled to scalar fields and construct a class of stable extremal black holes that are non-BPS, but still can be described by first-order differential equations. 
	These are driven by a ``superpotential'', which replaces the central charge $\cal Z$ in the usual black hole potential. 
	We provide a general procedure for finding this class and deriving the associated ``superpotential''. 
	We also identify some other cases which do not belong to this class, but show a similar behaviour. 
\end{titlepage}
\baselineskip 6 mm

\section{Introduction}

The ``no-hair theorem'' states that a black hole solution is completely specified by its mass $M$ and charges $Q$ (listing the angular momentum among the charges). 
Although there are by now several counter examples to a general validity of this theorem, it seems to hold true for spherically symmetric, static and asymptotically flat black holes in 4 dimensional gravity theories coupled to Maxwell fields. 
The same mass and charge are used to determine whether or not the black hole singularity is hidden by a horizon. 
This happens any time the mass is bigger or equal to the charge $M\geq |Q|$, providing a (sort of) BPS bound, which is saturated by extremal (zero temperature) black holes, having $M=|Q|$.

The idea of an attractor mechanism \cite{Ferrara:1995ih,Ferrara:1996dd,Ferrara:1996um} for supergravity black holes rests on the above arguments, leading to scalar fields that are drawn to fixed values, where they are functions only of masses and charges. 
More precisely, in Einstein-Maxwell theories coupled to scalar fields, the near horizon geometry of extremal black holes should depend only on the charges and not on the asymptotic values of the scalar fields.

In a supergravity, theory it is natural to expect that the charge giving the BPS bound be a central charge of the theory $Q = {\cal Z}$ (actually, the maximal eigenvalue of the central charges in extended supergravities \cite{Andrianopoli:1996ve}). 
The extremality condition for a given black hole is then equivalent to the requirement that some fraction of supersymmetry be preserved: the bound for the existence of a horizon can be identified with the BPS bound following from the supersymmetry algebra $M \geq |{\cal Z}|$. 
This is the context where the attractor mechanism was first introduced in \cite{Ferrara:1995ih,Ferrara:1996dd,Ferrara:1996um}.

In the beginning, the main attention was devoted to supersymmetric solutions because supersymmetry preserves the BPS bound at all values of the string coupling constant, and this in turn allows the analysis of the black hole entropy from a string theory point of view. 
However it is known that there exist also non-BPS extremal black holes, and that an attractor behaviour can be found also for these solutions, when perturbatively stable \cite{Goldstein:2005hq,Tripathy:2005qp}. 
Indeed, the attractor mechanism seems to be related to the extremality rather than to the supersymmetry property of a given solution \cite{Ferrara:1997tw,Tripathy:2005qp}. 
Despite some common features with the BPS case, the non-BPS extremal black holes are not expected to share the property of fulfilling first-order rather than second order Einstein and scalar field equations, a feature that arises as a consequence of the supersymmetry transformations on the fermions. 
However, it seems natural to ask whether they satisfy only second order differential equations or if there is also some first-order formalism which identically solves the equations of motion, and that could be related to the appearance of an attractor behaviour. 
If such a formalism would exist, it should be similar to supersymmetry, though different in some essential details. 
In the case of domain-wall solutions, this issue has been addressed and solved by ``fake supergravities'' \cite{Freedman:2003ax,Celi:2004st,Zagermann:2004ac,Skenderis:2006jq}. 
These are gravitational theories in d-dimensions that, in spite of not being supersymmetric in general, present some ``fake BPS equations'' for the metric and scalar fields that are of first order and originate from the vanishing action of certain operators on spinor parameters. 
Indeed, fake supergravity allows to construct stable domain-wall solutions satisfying first-order, attractor-like equations that are not following from supersymmetry. 
Spherically symmetric, static and asymptotically flat black hole solutions can be reduced to a one-dimensional problem of evolution in a radial coordinate, which resembles very closely the description of domain-walls. 
It is therefore interesting to see whether the analogy extends any further, also in the light of similar investigations on developing a first order formalism for cosmological solutions \cite{Bazeia:2005tj,Skenderis:2006fb}.

Building on previous knowledge of flow equations and other aspects of domain wall supergravity solutions \cite{Skenderis:1999mm,Ceresole:2001wi,Ceresole:2006iq}, in this note we address this problem by looking for non-BPS extremal black holes satisfying first-order equations. 
Although we will not give a general answer to the question of whether all extremal solutions derive from first-order equations, we are going to show that there exist classes of non-BPS extremal black holes of this type, and provide the conditions required to obtain them. 
Our solutions fulfill ordinary supergravity equations of motion as well as a fake supergravity first-order formalism.

We work for convenience in ${\cal N}=2$ supergravity in four dimensions. 
In this context, the superpotential $W(\phi)$ yielding BPS black holes is to be identified with the covariantly holomorphic central charge ${\cal Z}(\phi)$, that specifies the BPS solutions. 
In fact, the warp factor and the scalar field derivatives are related to ${\cal Z}(\phi)$ and its first-order derivative by the supersymmetry conditions. 
In the class of extremal solutions we are going to present, it is not ${\cal Z}(\phi)$ that appears in the BPS equations, but rather we find it replaced by another function $W(\phi)$ playing the role of a ``fake superpotential'' .

We start in section 2 with some general remarks and we build the set up to compare black hole and domain wall solutions. 
We will find that the non-extremality parameter $c^2\neq 0$ is related to the presence of a \textit{positive} curvature $\Lambda>0$ on the domain-wall. 
Then, in section 3 we will show how one can construct a general class of non-BPS extremal black holes by analyzing the symmetry properties of the supergravity potentials $V(\phi)$, and giving evidence of their degenerate description in terms of a superpotential $W(\phi)$. 
This construction was inspired by the suggestion that there may exist a canonical transformation relating BPS and non-BPS black holes at the horizon \cite{Kallosh:2006bx}.

In section 4 we show that this class is not empty by providing a simple example. 
Although the conditions that select this class of solutions are quite restrictive and difficult to match, this same construction can also be useful when the conditions are met only in some truncated setups. 
For this reason, always in section 4, we provide an example of non-BPS extremal black holes satisfying first-order differential equations in the STU model. 
In section 5 we give yet another example of non-BPS extremal solutions that do not belong to this class but show the same behaviour. 
In section 6, we end with some open problems.

\section{Black Holes and Domain Walls}

As explained in the introduction, we consider four-dimensional Einstein-Maxwell theories coupled to $n$ complex scalar fields $z^i$, with lagrangian 
\begin{equation}
	{\cal L} = -\frac{R}{2} + g_{i\bar\jmath} \partial_\mu z^i \partial_\nu \bar z^{\bar \jmath} + \hbox{Im}{\cal N}_ {\Lambda\Sigma} F^\Lambda_{\mu \nu} F^{\Sigma\,\mu\nu} +\hbox{Re}{\cal N}_{\Lambda\Sigma}F^\Lambda_{\mu\nu}(*F^\Lambda)^ {\mu\nu}. 
	\label{StartingAction} 
\end{equation}
The vector kinetic matrix ${\cal N}_{\Lambda\Sigma}(z,\bar z)$ is a complex and symmetric function of the scalar fields, $\Lambda=0, 1,\ldots , n_V$. 
We focus on black hole solutions of this system, especially concentrating on spherically symmetric, charged, static and asymptotically flat solutions, as they are known to display an attractor behaviour. 
For these reasons, the metric Ansatz is 
\begin{equation}
	ds^2 = - {\rm e}^{2U(r)} dt^2 + {\rm e}^{-2U(r)} \left[c^4 \frac{dr^2}{\sinh^4 (cr)} + \frac{c^2}{\sinh^2 (cr)} \left(d\theta^2 + \sin^2 \theta \, d\phi^2\right)\right]. 
	\label{ansatz} 
\end{equation}
We allow the scalar fields to have a profile in the radial direction, but fix the vector fields so that their field-strengths obey the usual quantization conditions 
\begin{equation}
	\int_{S^2} F^\Lambda = 4 \pi p^\Lambda, \qquad \int_{S^2} {\cal G}_\Lambda = 4 \pi q_\Lambda, \label{FG} 
\end{equation}
where $q_\Lambda$ and $p^\Lambda$ are the electric and magnetic charges respectively.

Since we are looking at time-independent solutions that preserve spherical symmetry, we can reduce the 4-dimensional action to a one-dimensional effective theory, by integrating over ${\mathbb R}_t \times S^2$ and discarding (infinite) constant integration factors. 
The resulting effective action is given by integrating over the remaining radial coordinate $S = \int dr {\cal L} $ the Lagrangian \cite{Ferrara:1997tw} 
\begin{equation}
	{\cal L} = (U^\prime(r))^2 + g_{i\bar \jmath} z^{\prime i}\bar z^{\prime \bar \jmath} + {\rm e}^{2U} V_{BH} - c^2, \label{BHeff} 
\end{equation}
with the prime denoting the derivative with respect to the radial coordinate (obviously one could discard the last constant term, but we will keep it for comparison with the domain-wall action). 
The first and the last term in (\ref{BHeff}) come from the Einstein-Hilbert action, while the derivatives on the scalar fields arise from their kinetic term, and the black hole effective potential $V_{BH}$ comes from the vector field terms and it is positive semi-definite. 
This effective action is actually quite general for any 4-dimensional gravity theory, provided the effective potential is tuned with the theory under consideration.

The original theory (\ref{StartingAction}) gives rise to some equations of motion that coincide with those of the above effective theory only up to a Hamiltonian constraint: 
\begin{equation}
	(U^\prime(r))^2 + g_{i\bar \jmath} z^{\prime i}\bar z^{\prime \bar \jmath} = {\rm e}^{2U} V_{BH} + c^2. 
	\label{constraint} 
\end{equation}
Therefore, black holes are solutions to the equations of motion for the lagrangian (\ref{BHeff}) 
\begin{eqnarray}
	U'' & = & {\rm e}^{2U} V_{BH}, \label{Upp}\\[2mm]
	z''{^i} + \Gamma_{jk}^i z^{\prime j}z^{\prime k} &=& {\rm e}^{2 U} g^{i\bar \jmath} \partial_{\bar \jmath} V_{BH}, \label{zpp} 
\end{eqnarray}
complemented by the Hamiltonian constraint (\ref{constraint}), where we can now identify $c^2 = 2 ST$ \cite{Ferrara:1997tw} for $S$ the entropy and $T$ the temperature of the black hole. 
Extremal black holes are those that have $c = 0$, following from their zero temperature $T=0$.

Let us now analyze the relation between this black hole effective action and the one for domain-walls in 4 dimensions.

It was already pointed out in \cite{Behrndt:2001qa} that supersymmetric black holes arise from an effective action that can be related to the one of flat supersymmetric domain-wall solutions. 
It was also suggested that the negative curvature $\Lambda < 0$ of the supersymmetric domain-wall solutions could be related to the angular momentum of supersymmetric black holes \cite{Behrndt:2001mx}. 
Here we want to examine this analogy for any extremal (also non-BPS) black hole as well as for non-extremal solutions, and we want to compare them to curved domain wall solutions. 
We will find that a \textit{positive} curvature $\Lambda >0$ of the domain-wall can be related to the non-extremality parameter of the black hole, namely $c^2$. 
We are led to this parallel because, as we will see in the following, both these constants play the role of deformation parameters in the first-order differential equations that describe the solutions.

In the case of domain-wall solutions in 4-dimensions, the metric Ansatz is 
\begin{equation}
	ds^2 = {\rm e}^{2 U(r)}\hat g_{ij} dx^i dx^j+ {\rm e}^{p U(r)}dr^2, \label{DWmetric} 
\end{equation}
for $p$ real, and where the 3-dimensional metric $\hat g$ is chosen among the following three possibilities: 
\begin{equation}
	\begin{array}{rcrcl}
		dS_3 & & \hat g_{ij} dx^i dx^j & = & -dt^2 + {\rm e}^{2 \sqrt{\Lambda} t}(dx_1^2 + dx_2^2)\qquad \Lambda > 0, \\[2mm]
		AdS_3 & & \hat g_{ij} dx^i dx^j & = & d\tau^2 + {\rm e}^{-2\sqrt{-\Lambda} \tau}(-dt^2 + dx^2) \qquad \Lambda < 0,\\[2mm]
		M_3 & & \hat g_{ij} dx^i dx^j & = & -dt^2 + dx_1^2 + dx_2^2, \qquad \Lambda = 0. 
	\end{array}
	\label{3dmetric} 
\end{equation}
When the domain-wall is supported by scalar fields (we neglect charged domain-walls like those constructed in \cite{Cacciatori:2002qx,Celi:2006pd}), and for $p=2$, the effective action follows from the lagrangian 
\begin{equation}
	{\cal L} = {\rm e}^{2 U(r)}\left[(U^\prime(r))^2 - g_{i\bar \jmath} z^{\prime i}\bar z^{\prime \bar \jmath} -{\rm e}^{2U} V_{DW} + \Lambda\right], \label{LagrangeDW} 
\end{equation}
with the constraint 
\begin{equation}
	(U^\prime(r))^2 -\frac13 g_{i\bar \jmath} z^{\prime i}\bar z^{\prime \bar \jmath} = - {\rm e}^{2U} V_{DW} + \Lambda. 
	\label{constDW} 
\end{equation}
The corresponding equations of motion for the warp factor and the scalar fields are 
\begin{eqnarray}
	U'' &=& - {\rm e}^{2 U} V_{DW} - \frac13 g_{i\bar \jmath} z^{\prime i}\bar z^{\prime \bar \jmath}, \label{UppDW} \\[2mm]
	z''{^i} + \Gamma_{jk}^i z^{\prime j}z^{\prime k} &=& U^\prime z'{^i} +{\rm e}^{2 U} g^{i\bar \jmath} \partial_{\bar \jmath} V_{DW}. 
	\label{zppDW} 
\end{eqnarray}

\subsection{First order equations}

For domain-wall solutions, we know since \cite{DeWolfe:1999cp,Freedman:2003ax} that whenever the scalar potential $V_{DW}$ is determined by a real superpotential $W$ such that 
\begin{equation}
	V_{DW} = - W^2 +\frac43 \frac{1}{\gamma^2} g^{i\bar \jmath} \partial_i W \partial_{\bar \jmath} W, \label{VDW} 
\end{equation}
the solution to the equations of motion coming from (\ref{LagrangeDW}) can also be derived from the first-order ``flow'' equations 
\begin{eqnarray}
	U^\prime &=& \pm {\rm e}^{U}\gamma(r) W, \label{UprimeDW}\\[2mm]
	z^{\prime i} &=& \mp {\rm e}^{U} \frac{2}{\gamma^2} g^{i\bar \jmath} \partial_{\bar \jmath} W, \label{phiprimeDW} 
\end{eqnarray}
where 
\begin{equation}
	\gamma \equiv \sqrt{1 + {\rm e}^{-2 U}\frac{\Lambda}{W^2}}. 
	\label{gammaDW} 
\end{equation}
It should be noted that the form of the potential (\ref{VDW}) is such that the constraint (\ref{constDW}) is identically satisfied upon using (\ref{UprimeDW}) and (\ref{phiprimeDW}).

At this stage, we are ready to explore the differences and similarities between the two setups. 
Firstly, the lagrangian (\ref{LagrangeDW}) can actually be related to the one in (\ref{BHeff}) by a conformal rescaling, provided that two more crucial sign changes in the scalar kinetic term and in the (cosmological) constant are taken into account. 
Clearly, these are precisely the two ingredients that give rise to different physical systems. 
However, the remaining constraints can be exactly mapped onto each other, at least for the case of constant scalar fields, when $V_{DW}$ is a negative constant, while $V_{BH}$ is a positive one. 
In this case (\ref{constDW}) matches (\ref{constraint}), once the identifications\ 
\begin{equation}
	V_{BH} = - V_{DW} = Q^2 \geq 0 \quad \hbox{ and }\quad c^2 = \Lambda \geq 0 \label{identifications} 
\end{equation}
are made (thus restricting to dS or Minkowski domain-walls). 
As we will see briefly, this feature implies a similar description of the two systems.

In particular, given this similarity, one can hope to reproduce also in the case of black holes the derivation leading to first-order equations for domain-walls, at least in simple theories where the scalar fields are constant.

As we have seen for the domain wall solutions, first order equations may be obtained by solving the constraint (\ref{constraint}), and then checking that the resulting dynamical flows also fulfill the equations of motion. 
For constant scalars, this procedure together with the identifications suggested above, gives immediately the result obtained in \cite{Lu:2003iv,Miller:2006ay}, for non-extremal black hole solutions with a single magnetic charge. 
Indeed, for a single magnetic charge $Q$, the black hole potential is $V_{BH} = Q^2$. 
Using the identifications $W = Q$ and $\Lambda = c^2$, we can solve the constraint (\ref{constraint}) by using 
\begin{equation}
	U^\prime = {\rm e}^{U}\gamma(r) W, \label{UpBH} 
\end{equation}
where 
\begin{equation}
	\gamma \equiv \sqrt{1 + {\rm e}^{-2 U}\frac{c^2}{Q^2}}. 
	\label{gammaBH} 
\end{equation}
Since the scalars are constant, equation (\ref{zppDW}) is identically satisfied and the warp factor equation of motion (\ref{UppDW}) (using again the identification $V_{BH} = - V_{DW}$) is equivalent to (\ref{Upp}). 
As we have just stated, this equation is implied by (\ref{UprimeDW}), and we can safely argue that (\ref{UpBH}) solves both the black hole equations of motion (\ref{Upp}) and the constraint (\ref{constraint}). 
This first-order equation for the warp factor is precisely the one derived in \cite{Lu:2003iv,Miller:2006ay}, expressed in our coordinate basis.

It is interesting to point out that this similarity implies a relation between the non-extremality parameter $c^2$, which drives the black hole away from the BPS condition, and the positive cosmological constant $\Lambda$, which in the domain-wall case forbids supersymmetric solutions\footnote{It is known that there are exceptions to this rule, when the total metric describes four-dimensional Minkowski or Anti de Sitter space. 
In this case, the de Sitter domain walls can be foliations of supersymmetric spaces, such that the supersymmetry parameters are not preserved on each foil.}. 
The final result is that a de Sitter curved domain wall in a gauged supergravity theory and a non extremal black hole in a Maxwell + Einstein theory with a single charge from an abelian vector field and no scalar fields (for instance pure supergravity with a non-trivial graviphoton charge) share the same effective action.

As we have noted before, it is not to be expected that the actions describing the domain-wall and black hole systems can be generically mapped exactly one on the other. 
The difference in the two systems shows up when trying to extend the first-order formalism just discussed to the case of non-constant scalar fields. 
While a warp-factor equation like the one suggested in (\ref{UpBH}) may solve the related equation of motion, there is no simple way to find a first-order equation with a similar property also for the scalar fields. 
The only instance where this seems to be possible is the case of \textit{extremal} black holes, where $c = 0$. 
For vanishing extremality parameter, the constraint (\ref{constraint}) becomes a relation between the potential, the derivatives of the scalar fields and the warp factor. 
Therefore, it becomes easy to see that if the constraint (\ref{constraint}) can be solved by a real ``superpotential'' function $W(z, \bar z)$ such that 
\begin{eqnarray}
	U^\prime &=& \pm {\rm e}^U W, \label{Uprime} \\[2mm]
	z^{\prime i} &=& \pm 2 {\rm e}^U g^{i\bar \jmath} \partial_{\bar \jmath} W, \label{zprime} 
\end{eqnarray}
then the potential $V_{BH}$ becomes 
\begin{equation}
	V_{BH} = W^2 + 4 g^{i\bar \jmath} \partial_i W \partial_{\bar \jmath}W\,, \label{VW} 
\end{equation}
and the equations of motion are identically satisfied. 
This can also be seen directly by rewriting the effective action (\ref{BHeff}) in the usual BPS form 
\begin{equation}
	S = \int dr \left[\left(U'\pm {\rm e}^U W\right)^2 + \left|z^{i\prime}\pm 2 {\rm e}^U g^{i\bar \jmath} \partial_{\bar \jmath}W\right|^2 \mp 2\frac{d}{dr}\left( {\rm e}^U W\right)\right]. 
	\label{BPSform} 
\end{equation}

The parallelism with the domain-wall solutions is self-manifest. 
The $c = 0$ constraint gives a situation similar to that of flat domain-walls. 
The first-order equations (\ref{UprimeDW}) and (\ref{phiprimeDW}) as well as the potential (\ref{VDW}) easily reduce to expressions which closely resemble the black hole ones (\ref{Uprime})-(\ref{zprime}) and (\ref{VW}), with an appropriate sign change in the scalar equation, reflecting the different relative sign (and factor) in the scalar potential. 
We stress once more that we need the solution to be extremal in order to solve the Hamiltonian constraint and therefore to implement first-order equations of motion when the scalars are not constant. 
The constraint for non-extremal black holes is shifted by the constant $c$ and it cannot be put easily in the appropriate first-order form that also satisfies the equations of motion. 
We also remark that nothing has been related to supersymmetry so far. 
On the other hand, it is clear that a potential of the form (\ref{VW}) is not common to any gravity theory.

In supersymmetric theories there is a natural superpotential function, which is defined by the central charge ${\cal Z}$. 
For instance, in ${\cal N} = 2$ supergravity the effective potential can be written in terms of ${\cal Z}$ as 
\begin{equation}
	V_{BH} = |{\cal Z}|^2 + g^{i \bar \jmath}D_i {\cal Z}D_{\bar \jmath} \overline{\cal Z}, \label{potential} 
\end{equation}
where $D_i= \partial_i+ \frac{1}{2} \partial_i K$ are K\"ahler covariant derivatives. 
This potential can be compared to (\ref{VW}) by identifying $W = |{\cal Z}|$ in which case (\ref{potential}) becomes 
\begin{equation}
	V_{BH} = |{\cal Z}|^2 + 4 g^{i \bar \jmath} \partial_i |{\cal Z}| \partial_{\bar \jmath} |{\cal Z}|, \label{pot2} 
\end{equation}
where we have used that the central charge is a covariantly holomorphic function of the scalar fields, i.e.~satisfying $\left(\bar \partial_{\bar \imath}-\frac12 \partial_{\bar \imath} K \right){\cal Z}=0$, which implies that ${\cal Z} = {\rm e}^{K(z^i,\bar z^{\bar \jmath})/2} {f}(z^i)$. 
Extrema of this superpotential function, with $D_i{\cal Z}=0$ and ${\cal Z} \neq 0$ (which therefore are in one to one correspondence with $ \partial_i |{\cal Z}| = 0$), give rise to supersymmetric black holes, and the first-order equations (\ref{Uprime})--(\ref{zprime}) are nothing but the conditions following from the (vanishing of the) supersymmetry variations of the supergravity fermi fields\footnote{Often in the literature the BPS equation for the scalar fields contains the covariant derivative on ${\cal Z}$ instead of the simple derivative on the absolute value $|{\cal Z}|$. 
The two equations are actually equivalent upon using the condition on the phase of the central charge following from the supersymmetry condition coming from the gravitino transformation in the radial direction $\delta \psi_r^A =0$. 
This is also clear from the BPS form of the action (\ref{BPSform}), which, for $W = |{\cal Z}|$, can be rewritten in the form given in \cite{Ferrara:1995ih,Ferrara:1997tw} using special geometry.} \cite{Ferrara:1995ih,Ferrara:1997tw}. 
However, we are interested in those theories that may not be supersymmetric, or in those supersymmetric theories where the constraint (\ref{constraint}) admits multiple solutions. 
This may be due to a potential (\ref{VW}) that does not univocally identify a superpotential $W$, but rather may be equivalently rewritten in terms of different superpotentials $W$, only one of which correspond to the true central charge $\cal Z$. 
Then, the first-order equations do not imply anymore preserved supersymmetries, as they differ from the Killing spinor equation and susy rules. 
At most, we can talk about pseudo or ``fake supersymmetries'', and preservation of the form of the potential will still grant the stability of the solution, if non singular.

\subsection{Multiple $W$ for the same $V_{BH}$}

In view of the above discussion, we now turn to explore the conditions for the constraint (\ref{constraint}) to be solved by a ``fake'' black hole superpotential $W$, which is not simply proportional to the central charge $\cal Z$. 
When such a real function $W(z, {\bar z})$ exists, its critical points, ${ \partial_i} W=0$, give rise to stable non-BPS black holes.

First of all, it must be clear that quite generally there is no unique solution to the effective potential $V$ in terms of a superpotential $W$ that preserves the ``stability'' form \cite{Skenderis:1999mm} appearing for instance in (\ref{VDW}). 
That expression should rather be interpreted as a partial differential equation defining $W$ for a given $V$. 
In the case of domain walls and just one scalar field, the issue has been raised in \cite{Martelli:2001tu,Sonner:2005sj}, showing that it is possible to have families of solutions. 
Moreover, also the fake supergravities are an indirect manifestation of this ambiguity. 
For our purposes, it is useful to consider the rescaled potential ${\tt V} (U, z, {\bar z}) = {\rm e}^{2U} V (z, \bar z)$. 
Then, ${\tt V} $ can be written as the sum of squares of derivatives of $\verb W (U, z, {\bar z})\equiv {\rm e}^U W(z, {\bar z})$ with respect to a set of effective coordinates that include also the warp-factor $x^A = \{U, z^i, \bar z^{\bar \imath}\}$: 
\begin{equation}
	{\tt V} (x^A) = g^{AB} \partial_A {\tt W}(x) \partial_B {\tt W}(x), \label{norma} 
\end{equation}
where $g_{UU} = 1$, $g_{Ui} = 0$ and $g_{AB}$ is positive definite. 
This formula shows how it is \emph{possible} to have different superpotentials, or better gradients of the superpotential $ \partial_A {\tt W}(x)$ generating the same $ {\tt V}(x)$.

The constraint (\ref{norma}) says that we will get the same {\tt V}$(x)$ for all the vectors $ \partial_A {\tt W}(x)$ having the same norm. 
Therefore, the constraint allows for a field dependent rotation, provided that the rotated vector be once more a gradient (at least locally). 
More in detail, given the same {\tt V}, we can write it in terms of two different superpotentials ${\tt W}$ and $\widetilde {\tt W}(x)$ provided 
\begin{equation}
	\partial_A {\tt W}(x) = R_A{}^B(z,{\bar z}) \partial_B {\widetilde {\tt W}}(x), \label{relationWWt} 
\end{equation}
and $R(z, \bar z)$ is a field-dependent rotation matrix $R^T g R = g$, which does not contain the warp factor, so that $\widetilde{\tt W} = {\rm e}^U \widetilde W(z,\bar z)$ In addition to the above contraints, this rotation matrix fulfills a differential condition: 
\begin{equation}
	d( dx^A R_A{}^B \partial_B \widetilde {\tt W}) = 0 \quad \Leftrightarrow\quad \partial_{[A}\left( R_{B]}{}^C \partial_C \widetilde {\tt W}\right) = 0. 
	\label{defrotation} 
\end{equation}
Although we have not found a general $R$ matrix to solve this condition, we will see in the next section that we can indirectly construct classes of solutions by using the symplectic structure of the 4-dimensional black hole potential.

Before proceeding, it is useful to pause and comment on some properties of the multiple choice of superpotentials to the same potential. 
The critical points of $V_{BH}$ with a finite area of the horizon necessarily have $ \partial_U {\tt W} \neq 0$, but $ \partial_i {\tt W = 0}$. 
This means that if we have two different superpotentials $W$ for the same potential $V_{BH}$, the generic behaviour of the transformation mapping them one into the other is such that critical points of one are mapped onto ordinary points of the other, with $ \partial_i \widetilde {\tt W} \neq 0$. 
Only critical points with zero value of the superpotential $W=0$ are common to all possible solutions of the potential constraint, but we exclude them because we want a non-trivial black hole horizon and this is related to $V_{BH} = W^2$ at the critical point. 
This further implies that when we solve the potential $V_{BH}$ for two different functions with only one of the two being the superpotential appearing in the supersymmetry transformations (i.e.~the central charge $\cal Z$ for $N=2$), the critical points of the other ``fake superpotential'' will describe horizons of non-supersymmetric black holes.

\section{One class of $V_{BH}$ with multiple $W$ }

From now on, we focus on ${\cal N } = 2$ supergravity, where the properties of special geometry allow us to specify a simple condition for finding multiple (fake) superpotentials describing the same black hole potential.

For this theory, the black hole potential is given by the invariant \cite{Ceresole:1995ca} 
\begin{equation}
	V_{BH}= I_1 = Q^T {\cal M} Q, \label{I1} 
\end{equation}
where $Q = \{p^\Lambda, q_\Lambda\} $ is the $Sp(2 n_V + 2, {\mathbb Z})$ symplectic vector of charges and ${\cal M}=(^{AB}_{CD})$ is the symplectic matrix defined by the entries 
\begin{equation}
	\begin{array}{rcl}
		A & = & -D^T = {\rm Re}{\cal N} ({\rm Im} {\cal N})^{-1},\\[2mm]
		C & = & ({\rm Im} {\cal N})^{-1}, \\[2mm]
		B & = & -{\rm Im} {\cal N} - {\rm Re} {\cal N}({\rm Im} {\cal N})^{-1}{\rm Re} {\cal N}. 
	\end{array}
	\label{entries} 
\end{equation}
This can be further rewritten through another symplectic matrix 
\begin{equation}
	M = \left( 
	\begin{array}{cc}
		D & C\\
		B & A 
	\end{array}
	\right) \label{calS} 
\end{equation}
via the relation 
\begin{equation}
	{\cal M} = {\cal I} M, \qquad {\cal I } = \left( 
	\begin{array}{cc}
		0 & -{\mathbb I}\\
		{\mathbb I} & 0 
	\end{array}
	\right), \label{rela} 
\end{equation}
with $M^2=-{\mathbb I}$. 
We remind that the matrix ${\cal N}_{\Lambda \Sigma}(z,\bar z)$, whose real and imaginary parts appear in the previous equation, defines the metric of the vector fields as in (\ref{StartingAction}).

As mentioned in the previous section, for ${\cal N} = 2$ supergravity this potential can always be written in terms of a superpotential, which is given by the central charge 
\begin{equation}
	{\cal Z } = {\rm e}^{K/2} \left(q_\Lambda X^\Lambda - p^\Lambda F_\Lambda\right) . 
	\label{Zord} 
\end{equation}
This is a symplectic invariant of the charge vector $Q = \{p^\Lambda, q_\Lambda\}$ and the covariantly holomorphic sections ${\cal V} ={\rm e}^{K/2}\{X^\Lambda, F_\Lambda\}\equiv \{L^{\Lambda},M_{\Lambda}\} $ (with $M_{\Lambda}={\cal N}_{\Lambda\Sigma}L^{\Sigma}$) describing the vector multiplet scalar manifold geometry: 
\begin{equation}
	{\cal Z } =Q^T {\cal I} {\cal V}=L^{\Lambda}q_{\Lambda}-M_{\Lambda}p^{\Lambda}. 
	\label{Zord2} 
\end{equation}
The black hole potential follows from (\ref{potential}) 
\begin{equation}
	V_{BH}=I_1= |{\cal Z}|^2 + g^{i\bar \jmath}D_i {\cal Z} D_{\bar\jmath} {\cal Z}, \label{I1Z} 
\end{equation}
and we have seen that this has a natural superpotential defined by $W = |{\cal Z}|$.

We first look at (\ref{rela}) and see that, given a potential defined by $I_1$, we still have the freedom to perform transformations on the charge vector $Q \to S Q$ without changing its value. 
Assuming $S$ is also a symplectic matrix, this happens when 
\begin{equation}
	I_1 = Q^T {\cal M} Q = Q^T S^T {\cal M} S Q \Rightarrow S^T {\cal M} S = {\cal M}. 
	\label{inv} 
\end{equation}
This last condition becomes 
\begin{equation}
	S^T {\cal I} M S = {\cal I} M, \label{cond} 
\end{equation}
and using the properties of symplectic matrices (and hence $S^T {\cal I} S = {\cal I} \Rightarrow S^T {\cal I} = {\cal I }S^{-1}$) we get 
\begin{equation}
	S^T {\cal I} M S = {\cal I }S^{-1} M S = {\cal I } M \label{boh} 
\end{equation}
and finally 
\begin{equation}
	[S,M] = 0. 
	\label{finale} 
\end{equation}
However, by the comparison of the two equivalent definitions (\ref{I1}) and (\ref{I1Z}) we deduce that only if $S$ is constant we can define a new ``fake superpotential'' 
\begin{equation}
	{\cal W} = Q^T S^T {\cal I} {\cal V} \label{WN2} 
\end{equation}
(with $W = |{\cal W}|$) giving rise to the same potential as ${\cal Z}$. 
Only in this case the derivatives on the superpotential will go through the matrix $S$ and therefore the charges transformed by the matrix $S$ will factorize, just like in the ordinary case where the superpotential is defined by the central charge, reconstructing ${\cal M}$.

Summing up, anytime we find a constant symplectic matrix $S$ that commutes with $M$ defined above, we find a new ``fake superpotential" whose critical points (if any) describe non-supersymmetric black holes.

Some comments are in order. 
Firstly, in the generic case the field dependent matrix $M$ spans all possible symplectic matrices in the group of duality transformations and therefore there will be no way to find constant matrices commuting with it. 
However, we will see that there is at least one simple instance where such matrix exists. 
Secondly, quite often one can consistently truncate the theory to a subset of the scalar fields, so that the constraint (\ref{finale}) allows for solutions. 
We will see this again in the next section for the case of $STU$ black holes. 
Finally, another strong constraint on the existence of such solutions comes from an observation in \cite{Ferrara:1997tw}, that the second derivatives of a potential of the form (\ref{potential}) at the critical point are proportional to the metric (and hence positive definite). 
This implies that matrices satisfying (\ref{finale}) can be found only if the Hessian of the black hole potential is positive-definite and proportional to the scalar $\sigma$-model metric also at the critical points describing the non-BPS black holes. 
Although there are examples where this happens, this cannot be a general feature of all non-BPS black holes. 
This does not exclude that there may be alternative ways of obtaining ``fake superpotentials'' describing the same potential. 
The main requirement then would be that these ``fake superpotentials'' must not be covariantly holomorphic functions of the scalar fields. 
In this case the argument in \cite{Ferrara:1997tw} fails and therefore there is no constraint on the Hessian of the potential at the non-BPS critical points. 
This is precisely the case where one can find a field dependent $R$ satisfying (\ref{defrotation}), but not a constant $S$ satisfying (\ref{finale}). 
We will see an example of this sort in section 5.

\section{Examples}

Let us now turn to examples of models where the black hole effective potential has multiple descriptions, by a superpotential ${\cal Z}$ and a ``fake superpotential'' $W$.

\subsection{One-modulus case}

The first example is given by the $SU(1,1)/U(1)$ model with just one modulus, generated by the prepotential 
\begin{equation}
	F = -i X^0 X^1. 
	\label{prep} 
\end{equation}
In special geometry, the K\"ahler potential is defined in terms of the sections $\{X^\Lambda,F_\Lambda\}$, with $F_\Lambda= \partial_\Lambda F(X)$, by 
\begin{equation}
	K = -\log[i(\bar X^\Lambda F_\Lambda - X^\Lambda \bar F_\Lambda)] , \label{KP} 
\end{equation}
and upon inserting (\ref{prep}) it reads $K = -\log 2(z+\bar z)$.

Using normal coordinates $z=X^1/X^0$, in the gauge $X^0 =1$ and for generic electric $q_\Lambda$ and magnetic $p^\Lambda$ charges, this generates a central charge: 
\begin{equation}
	{\cal Z} = \frac{q_0+i p^1 + (q_1 + i p^0) z}{\sqrt{2(z + \bar z)}} . 
	\label{Zcharge} 
\end{equation}
The black hole potential $V_{BH}$ is derived by inserting this expression in (\ref{potential}): 
\begin{equation}
	V_{BH} = \frac{(p^1)^2-i {q_1} ({z}-{\bar z}) {p^1}+{q_0}^2+i {p^0} {q_0} ({z}-{\bar z})+\left((p^0)^2+(q_1)^2\right) {z} {\bar z}}{{z}+{\bar z}}. 
	\label{potenziale1modulo} 
\end{equation}
Black hole solutions are then found by looking for solutions interpolating between flat space at infinity and $AdS_2 \times S^2$ at the horizons defined by the critical points of $V_{BH}$.

Critical points of $V_{BH}$ are found for 
\begin{equation}
	z^{\pm} = \frac{\pm (p^0p^1+q_0q_1)+i(p^0 q_0 - p^1 q_1)}{(p^0)^2 + (q_1)^2} , \label{zcrit} 
\end{equation}
and since Re$z > 0$, they lie inside the moduli space for $(p^0p^1+q_0q_1) >0$ when $z^+$ is chosen in (\ref{zcrit}), and $(p^0p^1+q_0q_1)<0$ for $z^-$.

This model has both supersymmetric black holes as well as non-supersymmetric ones. 
More precisely, $z^+$ (\ref{zcrit}) gives the supersymmetric vacuum, which satisfies $D_i {\cal Z} = 0$, with ${\cal Z}\neq 0$, (hence $ \partial_i |{\cal Z}| = 0$) and thus it is a fixed point of (\ref{Zcharge}) .

The negative sign $z^-$ gives the non-BPS black hole, for which $D_i {\cal Z} \neq 0$. 
The Hessian at these points is always positive as there are 2 identical positive eigenvalues 
\begin{equation}
	\pm\frac{1}{p^0p^1+q_0q_1}\{((p^0)^2 + (q_1)^2)^2,((p^0)^2 + (q_1)^2)^2\} . 
	\label{eigenH} 
\end{equation}

A simple inspection of the above formulae shows that the two type of black holes are related by a change of sign in the electric or magnetic charges. 
Let us now argue that this is precisely the transformation that can be achieved by acting on the charges with a matrix $S$, satisfying (\ref{finale}), to map the superpotential into the ``fake superpotential'' describing the non supersymmetric critical point and giving the first-order equations for the non supersymmetric black hole. 
As explained in the previous section, the black hole potential can be deduced from the (\ref{I1}) formula, by using the symplectic matrix (\ref{calS}), which reads 
\begin{equation}
	M = \left( 
	\begin{array}{cccc}
		\frac{y}{x} & 0 & -\frac{1}{x} & 0 \\[2mm]
		0 & -\frac{y}{x} & 0 & -\frac{(x)^2+(y)^2}{x} \\[2mm]
		\frac{(x)^2+(y)^2}{x} & 0 & -\frac{y}{x} & 0 \\[2mm]
		0 & \frac{1}{x} & 0 & \frac{y}{x} 
	\end{array}
	\right) \label{matrS} 
\end{equation}
where we have used the notation $z \equiv x + i y$.

We now look for a general constant symplectic matrix commuting with (\ref{matrS}). 
Given the simple structure of $M$, we can see that 
\begin{equation}
	S = -\cos[\theta] \left( 
	\begin{array}{cc}
		\sigma^3& 0\\[2mm]
		0 & \sigma^3 
	\end{array}
	\right) +\sin[\theta] \left( 
	\begin{array}{cc}
		0& -i \sigma^2\\[2mm]
		i \sigma^2 & 0 
	\end{array}
	\right) \label{Sex1} 
\end{equation}
is appropriate for this purpose. 
Applying this matrix to the charges we get a new \textit{complex} ``fake superpotential'' 
\begin{equation}
	{\cal W}= {\rm e}^{i \theta}\frac{-q_0+i p^1 + (q_1 - i p^0) z}{\sqrt{2(z + \bar z)}} \label{Zp} 
\end{equation}
that indeed differs from (\ref{Zcharge}), but gives rise to the same potential $V_{BH}$. 
It is also quite simple to check that the critical point of this new ``fake superpotential'' is the non-BPS black hole, namely (\ref{zcrit}) with the minus sign. 
Also, for $\theta =0$, this is proving what we were expecting: the BPS and non-BPS black hole are related by a sign change in the charges appearing in the definition of the superpotential. 
\begin{figure}
	[hbt] \centerline{ 
	\includegraphics[scale=1]{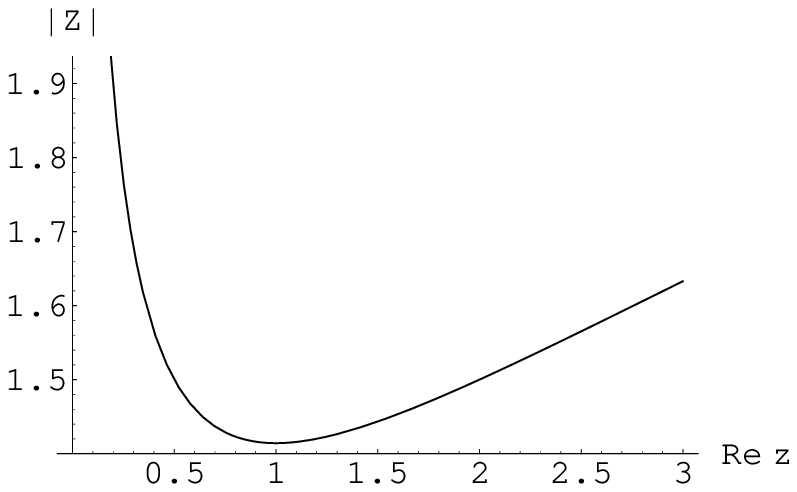} 
	\includegraphics[scale=1]{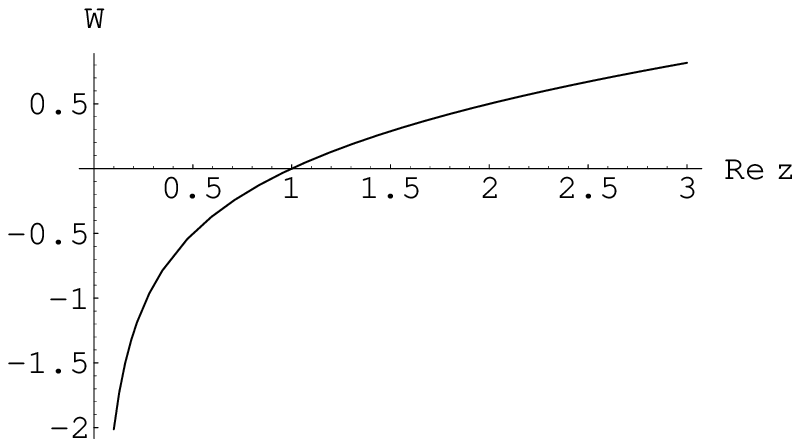} } \caption{Plots of the sections of $W$ and $|{\cal Z}|$ at Im $z=0$, for unit charges. 
	Where the central charge shows a minimum, the ``fake superpotentials'' crosses zero. 
	Changing the signs of the $q_0$ and $p^0$ charges exchanges the two pictures.} \label{plots} 
\end{figure}

From the general discussion of section 2, we also know that the superpotentials giving the first-order equations describing the two types of black holes, namely ${\tt W} = {\rm e}^U |{\cal Z}|$, for the supersymmetric one, and $\widetilde{\tt W} = {\rm e}^U |{\cal W}|$, for the non supersymmetric one, must be related by (\ref{relationWWt}). 
This is indeed what happens for a special choice of the matrix $R$, which is such that the norm of the superpotential and the norm of its derivatives with respect to the moduli fields are exchanged. 
This is a clear invariance of the black hole potential, which is given by the sum of the two, but it cannot always be realized with a rotation matrix satisfying (\ref{defrotation}) with the exception of the case of a single real scalar. 
The general way to construct a matrix that gives the desired result (exchange of $W^2$ with $| \partial W|^2$) is to take 
\begin{equation}
	R = \left( 
	\begin{array}{cc}
		0 & u^T \\[2mm]
		u & A 
	\end{array}
	\right), \label{Rmatrix} 
\end{equation}
where (for $n$ scalars) the $n$-dimensional vector $u$ is defined as the unit-norm derivative of the original superpotential with respect to the scalar fields 
\begin{equation}
	\vec u = \frac{\vec{\partial}\, {\tt W}}{| \vec \partial\, {\tt W}|} \label{defu} 
\end{equation}
and the $n\times n$ matrix $A$ acting on the scalar field directions is constructed as $A = {\cal R} - {\cal R}u \, u^T$, for ${\cal R}$ a rotation matrix. 
The relation (\ref{relationWWt}) can be explicitly verified by applying the above expressions 
\begin{equation}
	\partial \widetilde{\tt W} = \left( 
	\begin{array}{c}
		|\vec{\partial}\, {\tt W}| \\
		{\tt W}\; \vec u 
	\end{array}
	\right) = R\; \partial {\tt W} =R\; \left( 
	\begin{array}{c}
		{\tt W} \\
		|\vec \partial\, {\tt W}| \;\vec u 
	\end{array}
	\right). 
	\label{transfo} 
\end{equation}
For the simple case at hand, one can also check that the new superpotential is indeed the norm of the derivative of the supersymmetric one $|{\cal W}| = | \partial |{\cal Z}||$ and that the constraint (\ref{defrotation}) is identically satisfied.

\subsection{STU black hole}

The STU model can be constructed starting from a cubic prepotential 
\begin{equation}
	F = \frac{X^1 X^2 X^3}{X^0}. 
	\label{FSTU} 
\end{equation}
Using special coordinates, the K\"ahler potential of the scalar $\sigma$-model reads 
\begin{equation}
	K = -\log \left(-i(z^1 - \bar z^1)(z^2 - \bar z^2)(z^3 - \bar z^3)\right) \label{KpotSTU} 
\end{equation}
and the central charge follows in the usual way from ${\cal Z} = {\rm e}^{K/2}(q_\Lambda X^\Lambda - p^\Lambda F_\Lambda)$. 
Let us focus on a choice of charges admitting both BPS and non-BPS black holes. 
This is the case when we have only one magnetic charge $p^0$ and 3 electric charges $q_i$, $i=1,2,3$. 
For this choice of charges, the central charge reads 
\begin{equation}
	{\cal Z} = {\rm e}^{K/2}\left(q_i z^i + p^0 z^1 z^2 z^3\right). 
	\label{ZSTU} 
\end{equation}
The BPS and non-BPS attractor points for the moduli are then given by 
\begin{equation}
	z^1 = -i \sqrt{\mp\frac{q_2 q_3}{p^0 q_1}} , \qquad z^2 = -i \sqrt{\mp\frac{q_3 q_1}{p^0 q_2}} , \qquad z^3 = -i \sqrt{\mp\frac{q_1 q_2}{p^0 q_3}} , \label{attra} 
\end{equation}
where the minus sign is for the BPS black hole ($p^0 q_1 q_2 q_3 <0$ in that case) and the plus sign is for the non supersymmetric ones ($p^0 q_1 q_2 q_3 >0$). 
These critical points are both stable, as shown in \cite{Kallosh:2006ib}. 
The non-BPS critical point, however, does not have a Hessian matrix for the potential which is positive definite, but there are flat directions in the axion sector. 
It is clear already from this fact that we should not expect to find a constant $S$ giving rise to a new ``fake superpotential'' generating this critical point. 
Indeed, by constructing explicitly $M$ for this example we do not find any constant $S$ commuting with it. 
If, on the other hand, we truncate the theory setting to zero the axion fields Re$z^i$, we can once more prove that there is a ``fake superpotential'' describing the non-BPS black hole.

First of all, we can prove that this truncation is consistent, as the axion equations of motion are identically satisfied by setting them to zero. 
We are therefore left with an effective potential depending only on Im$z^i = y^i<0$, and this latter follows from the truncated central charge 
\begin{equation}
	{\cal Z} = \frac{1}{\sqrt{8}}\left[-p^0 \sqrt{-y^1 y^2 y^3} + q_1 \sqrt{-\frac{y^1}{y^2 y^3}} + q_2 \sqrt{-\frac{y^2}{y^3 y^1}} + q_3 \sqrt{-\frac{y^3}{y^1 y^2}}\right], \label{Ztrunc} 
\end{equation}
whose critical point is (\ref{attra}), with the minus sign choice. 
The difference however is that now the matrix $M$ simplifies a lot 
\begin{equation}
	M = \left( 
	\begin{array}{cccccccc}
		0 & 0 & 0 & 0 & \frac{1}{y^1 y^2 y^3} & 0 & 0 &0 \\
		0 & 0 & 0 & 0 & 0 & \frac{y^1}{y^2 y^3} & 0 &0 \\
		0 & 0 & 0 & 0 & 0 & 0 & \frac{y^2}{y^3 y^1} &0 \\
		0 & 0 & 0 & 0 & 0 & 0 & 0 &\frac{y^3}{y^1 y^2} \\
		-{y^1 y^2 y^3} & 0 & 0 &0 & 0 & 0 & 0 &0 \\
		0 & -\frac{y^2 y^3}{y^1} & 0 &0 & 0 & 0 & 0 &0 \\
		0 & 0 & -\frac{y^3 y^1}{y^2} &0 & 0 & 0 & 0 &0 \\
		0 & 0 & 0 &-\frac{y^1 y^2}{y^3} & 0 & 0 & 0 &0 
	\end{array}
	\right) \label{ScalSTU} 
\end{equation}
and therefore we can find a constant $S$ commuting with it: 
\begin{equation}
	S = {\rm diag}\{a,b,c,d,a,b,c,d\} , \label{constS} 
\end{equation}
where $a,b,c,d= \pm 1$. 
For instance, choosing $a = -1$, $b=c=d=1$, we get the ``fake superpotential'' 
\begin{equation}
	W = \frac{1}{\sqrt{8}}\left[p^0 \sqrt{-y^1 y^2 y^3} + q_1 \sqrt{-\frac{y^1}{y^2 y^3}} + q_2 \sqrt{-\frac{y^2}{y^3 y^1}} + q_3 \sqrt{-\frac{y^3}{y^1 y^2}}\right], \label{fakeW} 
\end{equation}
which generates the same potential as (\ref{Ztrunc}), but the only critical point of $W$ is the non-BPS black hole with the plus sign in (\ref{attra}). 
Other choices of the signs in the matrix $S$ give ``fake superpotentials'' that do not admit critical points in the allowed regions of the moduli space. 
Once more, the relation between the two superpotentials is a sign change in the charges, but we will see in the next section that this is not a necessary condition in order to find multiple superpotential solutions to the potential constraint.

\section{More general solutions}

In this last section we consider another simple model with a single modulus. 
However, we choose a model that does not allow for solutions of (\ref{finale}), but still admits multiple black hole vacua. 
We will be able to identify also in this case multiple superpotentials driving the first-order equations yielding BPS and non-BPS black holes.

A model with these properties is the one with a cubic prepotential like the STU model, but with a single modulus and thus only two sections $X^\Lambda$: 
\begin{equation}
	F = \frac{(X^1)^3}{X^0}. 
	\label{prepbis} 
\end{equation}
Let us consider the case of black holes generated by the electric charge $q_1$ and the magnetic one $p^0$: 
\begin{equation}
	{\cal Z} = \frac{z q_1 +p^0 z^3}{\sqrt{-i(z - \bar z)^3}}. 
	\label{Zchargebis} 
\end{equation}
Critical points of the central charge lead to supersymmetric black holes at 
\begin{equation}
	z= -i \sqrt{-\frac{q_1}{3p^0}}, \label{zcritBPS} 
\end{equation}
provided $p^0 q_1 <0$. 
Analyzing the full black hole potential we can also find a non-supersymmetric critical point (and hence a non-BPS black hole horizon) at 
\begin{equation}
	z= -i \sqrt{\frac{q_1}{3p^0}}, \label{zcritnonBPS} 
\end{equation}
when $p^0 q_1 >0$. 
The Hessian at these points is always positive, but with different eigenvalues. 
This already implies that we cannot obtain a ``fake superpotential'' by acting on the charges in $\cal Z$ by a constant $S$ transformation. 
This can be explicitly verified by computing $M$ and checking that there are no constant matrices commuting with it. 
Still, we can find a ``fake superpotential'' generating (\ref{zcritnonBPS}) by making a field-dependent transformation as suggested in (\ref{relationWWt}). 
Following these considerations, the new superpotential cannot be covariantly holomorphic and this is indeed the case as one can see from the explicit expression, which reads: 
\begin{equation}
	W= \left|\frac{z q_1 +p^0 z^2 \bar z}{\sqrt{-i(z - \bar z)^3}}\right|. 
	\label{ZnonBPS} 
\end{equation}
This superpotential clearly differs from (\ref{Zchargebis}), but gives rise to the same potential $V_{BH}$ through (\ref{VW}). 
This new ``fake superpotential'' has a single critical point, which corresponds to the horizon of the non-BPS black hole, namely (\ref{zcritnonBPS}).

Finally, it is interesting to note that formally this same superpotential can be obtained by applying a field-dependent $S$ to the charges defining (\ref{Zchargebis}). 
This same $S$, once the axion is fixed at the critical point, reduces to a constant matrix that commutes with $M$ in the same truncated setup, just like in the $STU$ model.

\section{Outlook}

One of the main reasons why it is interesting that supersymmetric black hole solutions are described by first order equations is related to the possible existence of an attractor mechanism for the scalar fields. 
Henceforth, it is very appropriate to address this question in the class of examples we have just presented. 
The non-BPS black holes we have found in section 3 are for sure attractor points of the potential, as it is clear from the comments on the positivity of the Hessian of the scalar potential. 
However, there is more that we can say, and following the procedure used for the supersymmetric attractors in \cite{Ferrara:1996dd}, we can derive the attractor equations in an algebraic form.

In \cite{Ferrara:1996dd}, starting from the BPS extremality condition $D_i {\cal Z} = 0$ and using special geometry relations, one obtains a purely algebraic condition specifying the critical points 
\begin{equation}
	Q = \hbox{Im}\left({\cal Z}\overline{\cal V}\right) \label{oldattractor} 
\end{equation}
The same argument goes through in our case by replacing the central charge with the ``superpotential'' ${\cal W}$ and the charges with those transformed by the action of the matrix $S$. 
The final outcome is that for these examples the non-BPS critical points can be obtained by solving the algebraic equation 
\begin{equation}
	S Q = \hbox{Im} \left({\cal W} \overline{\cal V}\right). 
	\label{generalattractor} 
\end{equation}
In order to derive this result we had to use the holomorphicity properties of this ``superpotential'' and therefore we can not get a similar result in the more general case of a real $W$ generated by some $R$ transformation which is not related to the class just presented here. 
Both these classes will be described by the non-BPS attractor equations of \cite{Kallosh:2006bt}, though, and it would be desirable to reach a deeper understanding of the relation between the two formulations.

Although it is clear that our discussion has an easy extension to black hole solutions with a higher number of supersymmetries, we have worked for convenience in ${\cal N}=2$ supergravity in four dimensions. 
In this context, the superpotential yielding BPS black holes $W(\phi)$ is to be identified with the covariantly holomorphic central charge ${\cal Z}(\phi)$, that specifies the BPS solutions. 
In fact, the warp factor and the scalar field derivatives are related to ${\cal Z}(\phi)$ and its first-order derivative by the supersymmetry conditions. 
We have been able to generate new classes of extremal black hole solutions where the potential is not expressed in terms of the central charge ${\cal Z}(\phi)$, but rather in terms of the ``fake'' superpotential $W(\phi)$. 
It would be interesting to understand the physical meaning of this ``superpotential'' as some generalized (central) charge of the theory. 
This is especially true in virtue of the connection between the ``superpotential'' and the area of the non-BPS black hole horizon, that could be suggestive of some relation between this $W(\phi)$ and the entropy functional of \cite{Sen:2005wa}.

It is important to understand whether our class of solutions, or others that can be generated by a similar mechanism, are just special ones inside all non-BPS extremal black holes, or if there is a general way to argue for the same behaviour for all other extremal solutions.

Furthermore, the intriguing connection between canonical transformations relating BPS and non-BPS black holes at the horizon and the $S$ matrix transformation relating the central charge and the fake superpotentials certainly worths future investigations.

Finally, it would be very interesting to see whether the first order equations we provide in this note can be extracted from some pseudo supersymmetry transformation of a fully fledged fake supergravity, and we hope to report on it somewhere else.

\medskip 
\section*{Acknowledgments.}

\noindent We are glad to thank R.~D'Auria and especially M.~Zagermann for many useful comments. 
The research is supported by the European Union under the contract MRTN-CT-2004-005104, ``Constituents, Fundamental Forces and Symmetries of the Universe'' in which A.~C.~is associated to Torino University and G.~D.~is associated to Padova University.


\end{document}